\documentclass[showpacs,preprintnumbers,amsmath,amssymb]{revtex4}

\usepackage{graphicx}
\usepackage{dcolumn}
\usepackage{bm}
\usepackage{url}
\usepackage{epsfig}
\usepackage{multirow}
\usepackage{tensor}

\def\be{\begin{equation}}
\def\ee{\end{equation}}
\def\ba{\begin{eqnarray}}
\def\ea{\end{eqnarray}}

\newcommand{\fr}[2]{\frac{#1}{#2}}

\newcommand{\ob}{\omega_{b}}
\newcommand{\oa}{\omega_{a}}
\newcommand{\w}{\omega}
\newcommand{\Ob}{{\cal O}}

\newcommand{\rhode}{\rho_{\rm{DE}}}
\newcommand{\rhodeo}{\rho_{\rm{DE}0}}
\newcommand{\Odeo}{\Omega_{\rm{DE}0}}
\newcommand{\Omo}{\Omega_{\rm{m}0}}

\newcommand{\Fapp}{\widetilde{F}}
\newcommand{\PRD}{Phys.\ Rev.\ D}
\newcommand{\PRL}{Phys.\ Rev.\ Lett.}
\newcommand{\JCAP}{J.\ Cosmol.\ Astropart.\ Phys.}
\newcommand{\MNRAS}{Mon.\ Not.\ Roy.\ Astron.\ Soc.}

\begin{document}

\title{Optimal strategies : theoretical approaches to the parametrization of the dark energy equation of state}


\author{Seokcheon Lee}
\email{skylee@kias.re.kr}
\affiliation{School of Physics, Korea Institute for Advanced Study, Heogiro 85, Seoul 130-722, Korea}


\begin{abstract}
The absence of compelling theoretical model requires the parameterizing the dark energy to probe its properties. The parametrization of the equation of state of the dark energy is a common method. We explore the theoretical optimization of the parametrization based on the Fisher information matrix. As a suitable parametrization, it should be stable at high redshift and should produce the determinant of the Fisher matrix as large as possible. For the illustration, we propose one parametrization which can satisfy both criteria. By using the proper parametrization, we can improve the constraints on the dark energy even for the same data. We also show the weakness of the so-called principal component analysis method.

\end{abstract}


\maketitle

We consider strategies for the most accurate determination of the dark energy (DE) parameters using cosmological observables. Due to the absence of compelling theoretical model, the parameterizing the DE by its equation of state (eos) $\omega$ is commonly used in the analysis. It is usually considered what is the optimal redshift distribution to best constrain those parameters \cite{9805117, 0012510}. However, we contemplate the theoretical optimization by using the Fisher information matrix and compare the different parametrization of $\w$.

If one has a dataset of observable $\Ob_i$ with redshifts $z_{i}$, $i = 1, \cdots, N$, one may compute the least squares estimates of the cosmological parameters by minimizing \be \chi^2_{\Ob} = \sum_{i}^{N} \fr{\Bigl( \Ob_i - \Ob(z_i, \vec{p}) \Bigr)^2}{\sigma_i^2} \, , \label{chi2} \ee where $\Ob_i$ is the measured quantity, $\Ob (z_i, \vec{p})$ is the expected value of the observable for the redshift $z_i$, $\vec{p}$ is the set of cosmological parameters to estimate, and $\sigma_i$ is the error on the measurement. Then, the Fisher information matrix is defined to be \be F_{lm} = \Biggl< \fr{1}{2} \fr{\partial^2 \chi_{\Ob}^2}{\partial p_l \partial p_m} \Biggr>_{p_{l}^{\ast}, p_{m}^{\ast}} \, , \label{Fisher} \ee where $p_{l, \, m}^{\ast}$ are the parameter estimates ({\it i.e.} where $\chi_{\Ob}^2$ becomes minimum). Since the likelihood function is approximately Gaussian near the maximum likelihood (ML) point, the covariance matrix for a maximum likelihood estimator is given by \be \Bigl( C^{-1} \Bigr)_{lm} = \fr{1}{2} \fr{\partial^2 \chi_{\Ob}^2}{\partial p_l \partial p_m} \Biggl|_{p_{l}^{\ast}, p_{m}^{\ast}} \, . \label{InvC} \ee The Fisher information matrix is simply the expectation value of the inverse of the covariance matrix at the ML point. From now on, we will omit the square bracket and the evaluating parameter estimates in the Fisher matrix for the convenience.

The iso-$\Delta \chi_{\Ob}^2$ contour in the parameter space is approximated by the quadratic equation \be (\Delta \vec{p})^{T} F \Delta \vec{p} = \Delta \chi_{\Ob}^2 \, , \label{Deltachi2} \ee where $\Delta \vec{p} = \vec{p} - \vec{p}^{\ast}$ is the deviation of the parameters from the fiducial value and $F$ is the Fisher matrix given in Eq. (\ref{Fisher}). Thus, the most accurate determination of $n$ cosmological parameters is described by a minimal volume of the n-dimensional ellipsoid. The volume of the ellipsoid is given by \be V \propto \fr{1}{\sqrt{\det (F/\Delta \chi_{\Ob}^2)}} \, . \label{V} \ee Thus, the determination of $F$ should be maximum to minimize the volume of the uncertainty ellipsoid.

By using Eqs. (\ref{chi2}) and (\ref{Fisher}), we obtain \ba F_{lm} &=& - \sum_{i=1}^{N} \fr{\Ob_i - \Ob(z_i, \vec{p})}{\sigma_i^2} \fr{\partial^2 \Ob(z_i, \vec{p})}{\partial p_l \partial p_m} + \sum_{i=1}^{N} \fr{1}{\sigma_i^2} \fr{\partial \Ob(z_i, \vec{p})}{\partial p_{l}} \fr{\partial \Ob(z_i, \vec{p})}{\partial p_{m}} \nonumber \\ &\simeq& \sum_{i=1}^{N} \fr{1}{\sigma_i^2} \fr{\partial \Ob(z_i, \vec{p})}{\partial p_{l}} \fr{\partial \Ob(z_i, \vec{p})}{\partial p_{m}} \equiv \Fapp_{lm} \, , \label{Fisher2} \ea where we use $\sum \fr{\Ob_i - \Ob(z_i, \vec{p})}{\sigma_i^2} \fr{\partial^2 \Ob(z_i, \vec{p})}{\partial p_l \partial p_m} \simeq 0$ and define the approximate Fisher matrix $\Fapp_{lm}$ including only the second term in the original $F_{lm}$. We can use this approximation because $\chi^2$ becomes minimum at ML point $\vec{p}^{\ast}$ \be \fr{1}{2} \fr{\partial \chi^2}{\partial p_l} \Biggl|_{p_{l}^{\ast}} = - \sum_{i=1}^{N} \fr{\Ob_{i} - \Ob(z_i, \vec{p})}{\sigma_i^2} \fr{\partial \Ob(z_i, \vec{p})}{\partial p_l} \Biggl|_{p_{l}^{\ast}} = 0 \, . \label{chi2dp} \ee Eq. (\ref{chi2dp}) can be interpreted as the summation of $\fr{\partial \Ob(z_i, \vec{p})}{\partial p_l}$ weighted by $ \fr{\Ob_{i} - \Ob(z_i, \vec{p})}{\sigma_i^2}$ is zero. As long as the data converges to the fiducial values, the weighted summation of the variation of $\fr{\partial \Ob(z_i, \vec{p})}{\partial p_l}$ ({\it i.e.} $\sum \fr{\Ob_i - \Ob(z_i, \vec{p})}{\sigma_i^2} \fr{\partial^2 \Ob(z_i, \vec{p})}{\partial p_l \partial p_m}$) should be negligible. Thus, the first term of the right hand side of Eq. (\ref{Fisher2}) is much smaller than the second one. We will explicitly show this later in Table. \ref{table2} by using the currently available $H(z)$ data \cite{13035076, 0412269, 12013609, 08073551, 12043674, 09073149}.

In general, $\vec{p}$ includes the present value of the Hubble parameter $H_0$, the present value of the matter energy density contrast $\Omo$, the present value of the DE density contrast $\Odeo$ and the parameters of the eos of DE $\omega$. We concentrate on the flat universe and $\Odeo$ is determined as $1 - \Omo$. Now, we assume that the nuisance parameters $H_0$ and $\Omo$ are exactly known and only consider the estimation of the parameters of $\omega$. $\Fapp$ given in Eq. (\ref{Fisher2}) can be further specified depending on the observables. First, we consider when $\rhode$ is a direct variable of $\Ob(z_i, \vec{p})$. $H(z)$ corresponds to this case. Then, the approximate Fisher matrix becomes \ba \Fapp_{lm} &=& \sum_{i=1}^{N} \fr{1}{\sigma_i^2} \Biggl( 3 \fr{\partial \Ob (z_i, \vec{p})}{\partial \ln[ \rhode (z_i, \vec{p})]} \Biggr)^2 \Biggl( \int_{0}^{z_i} \fr{\partial \omega}{\partial p_{l}} d \ln (1+x) \Biggr) \Biggl( \int_{0}^{z_i} \fr{\partial \omega}{\partial p_{m}} d \ln (1+x) \Biggr) \, \label{Fisher2wH} \\ &\equiv& \sum_{i=1}^{N} G_{i}^2 \Biggl( \int_{0}^{z_i} \fr{\partial \omega}{\partial p_{l}} d \ln (1+x) \Biggr) \Biggl( \int_{0}^{z_i} \fr{\partial \omega}{\partial p_{m}} d \ln (1+x) \Biggr) \nonumber \, ,  \ea where we use $\rhode = \rhodeo \exp \Bigl[ 3 \int_{0}^{z} (1 + \omega(x)) d \ln(1+x) \Bigr]$. Second, if $\rhode$ is included in $\Ob$ as an integrand like in $d_{L}$, then $\Fapp$ is given by \ba \Fapp_{lm} &=& \sum_{i=1}^{N} \fr{1}{\sigma_i^2} \Biggl( \int_{0}^{z_i} \fr{3 \partial \Ob (x)}{\partial \ln[ \rhode(x)]} \Bigl[\int_{0}^{x} \fr{\partial \omega}{\partial p_{l}} d \ln (1+y) \Bigr] dx \Biggr) \Biggl( \int_{0}^{z_i} \fr{3 \partial \Ob (x)}{\partial \ln [\rhode(x)]} \Bigl[\int_{0}^{x} \fr{\partial \omega}{\partial p_{m}} d \ln (1+y) \Bigr] dx \Biggr) \, . \label{Fisher2wdL} \ea In general, we can express the approximate Fisher matrix as \ba \Fapp_{lm} &=& \sum_{i=1}^{N} \sum_{l=1}^{n} \sum_{m=1}^{n} G_{i}^2 W_{l}(z_i) W_{m}(z_i) \,\,\,\,\, ,  \mbox{where} \nonumber \\
G_i &=& \left\{ \begin{array}{ll}
\fr{H_0}{\sigma_i} \fr{3[(H(z_i)/H_0)^2 - \Omo(1+z_i)^3]}{2(H(z_i)/H_0)} & \mbox{if $\Ob = H(z)$} \\
\fr{1}{\sigma_i} & \mbox{if $\Ob = d_{L}(z)$} \end{array} \right. \,\,\,\,\,  \mbox{and} \label{Fisher3}  \\
W_l(z_i) &=& \left\{ \begin{array}{ll}
\int_{0}^{z_i} \fr{\partial \omega(x)}{\partial p_{m}} d \ln (1+x) & \mbox{if $\Ob = H(z)$} \\
\int_{0}^{z_i} \fr{3 \partial \Ob (x)}{\partial \ln[ \rhode(x)]} \Bigl[\int_{0}^{x} \fr{\partial \omega(y)}{\partial p_{l}} d \ln (1+y) \Bigr] dx & \mbox{if $\Ob = d_{L}(z)$}. \end{array} \right. \nonumber \ea
Thus, the determinant of the approximate Fisher matrix is given by \be \det(\Fapp) = \sum_{a,b,c, \cdots = 1}^{n} \varepsilon_{abc \cdots} \Bigl( \sum_{i=1}^{N} G_{i}^2 \Bigr)^n W_1W_{a} W_{2}W_{b} W_{3}W_{c} \cdots \, , \label{DetF} \ee where $\varepsilon_{abc \cdots}$, analogous to the Levi-Civita symbol, is $+1$ for even commutations $(abc \cdots)$ and $-1$ for odd commutations, and zero if any index is repeated.

First, we consider the models with two parameters $\omega = \omega_a + \omega_{b} f(z)$. In this case, $\det(\Fapp)$ is given by
\ba \det(\Fapp) &=& \left| \begin{array}{cc}
\sum G_{i}^2 \Bigl( W_{a}(z_i) \Bigr)^2 &
\sum G_{i}^2 \Bigl( W_{a}(z_i) \Bigr) \Bigl( W_{b}(z_i) \Bigr)  \\
\sum G_{i}^2 \Bigl( W_{b}(z_i) \Bigr) \Bigl( W_{a}(z_i) \Bigr)  &
\sum G_{i}^2 \Bigl( W_{b}(z_i) \Bigr)^2  \end{array} \right| \\
&=& \sum_{i=1}^{N-1} \sum_{j=i+1}^{N} G_{i}^2 G_{j}^2 \Biggl( \varepsilon_{ij} W_{a}(z_i) W_{b}(z_j) \Biggr)^2 \, . \label{detFw2} \ea If we choose the observable as $H(z)$, then $\det(\Fapp)$ becomes \ba \det(\Fapp)_{H} &\simeq& \sum_{i=1}^{N-1} \sum_{j=i+1}^{N} G_{i}^2 G_{j}^2 \Biggl( \ln[1+z_i] \int_{0}^{z_j} f(x) d \ln[1+x] - \ln[1+z_j] \int_{0}^{z_i} f(x) d \ln[1+x] \Biggr)^2 \nonumber  \\ &=& \sum_{i=1}^{N-1} \sum_{j=i+1}^{N} G_{i}^2 G_{j}^2 \left\{ \begin{array}{ll}
\Bigl( \ln[1+z_i] z_j - z_i \ln[1+z_j] \Bigr)^2 & \mbox{if $f(z) = z$} \\
\Bigl( \fr{1}{2} \ln[1+z_i] \ln[1+z_j] \ln \Bigl[\fr{1+z_j}{1+z_i} \Bigr] \Bigr)^2 & \mbox{if $f(z) = \ln[1+z]$} \\
\Bigl( \fr{z_i}{1+z_i} \ln[1+z_j] - \fr{z_j}{1+z_j} \ln[1+z_i] \Bigr)^2 & \mbox{if $f(z) = \fr{z}{1+z}$}
\end{array} \right. \, . \label{detFH} \ea We use the $H(z)$ data obtained from the Planck satellite mission \cite{13035076}, the passively evolving galaxies \cite{0412269, 09073149}, the differential spectroscopic evolution of early-type galaxies \cite{12013609}, the spectroscopic SDSS LRG galaxies \cite{08073551}, and Wiggle Z dark energy survey \cite{12043674}. This is shown in the table \ref{table1}.
\begin{center}
    \begin{table}
    \begin{tabular}{ | c || c | c | c | c | c | c | c | c | c | c | c | c | c |}
    \hline
    $z$  & 0  & 0.09  & 0.17 & 0.179  & 0.199 & 0.24 & 0.27 & 0.352 & 0.4 & 0.43 & 0.44 & 0.48 & 0.593 \\ \hline
    $H(z)$ & 67.4 & 69 & 83 & 75 & 75 & 79.69 & 77 & 83 & 95 & 86.45 & 82.6 & 97 & 104 \\ \hline
    $\sigma$ & 1.4 & 12 & 8 & 4 & 5 & 3.32 & 14 & 14 & 17 & 3.27 & 7.8 & 62 & 13 \\ \hline
    Ref & \cite{13035076} & \cite{0412269} & \cite{0412269} & \cite{12013609} & \cite{12013609} & \cite{08073551} & \cite{0412269} & \cite{12013609} & \cite{0412269} & \cite{08073551} & \cite{12043674} & \cite{09073149} & \cite{12013609} \\ \hline
    $z$  & 0.6  & 0.68  & 0.73 & 0.781  & 0.875 & 0.88 & 0.90 & 1.037 & 1.30 & 1.43 & 1.53 & 1.75 & \\ \hline
    $H(z)$ & 87.9 & 92 & 97.3 & 105 & 125 & 90 & 117 & 154 & 168 & 177 & 140 & 202 &  \\ \hline
    $\sigma$ & 6.1 & 8 & 7 & 12 & 17         & 40 & 23 & 20     & 17 & 18     & 14 & 40 &  \\ \hline
    Ref & \cite{12043674} & \cite{12013609} & \cite{12043674} & \cite{12013609} & \cite{12013609} & \cite{09073149} & \cite{0412269} & \cite{12013609} & \cite{0412269} & \cite{0412269} & \cite{0412269} & \cite{0412269} & \\ \hline
   \end{tabular}
    \caption{H(z) from currently available observation.}
    \label{table1}
    \end{table}
\end{center}
In Eq. (\ref{detFH}), we consider three well known parametrization \cite{0411803, 9904356, 0009008, 0208512}. This equation shows the importance of the parametrization of DE eos. The values of $G_{i}$ and $G_{j}$ should be very close to each other for the different models and thus the magnitude of $\det(\Fapp)$ depends only on the geometric part of $\omega$. Thus, the magnitude of $\det(\Fapp)$ depends on the magnitude of the quantities in the square bracket. It means that we can extract the better constraints depending on the parametrization of $\omega$ even for the same measurement. $f(z) = z$ or $\ln[1+z]$ gives the stronger constraints on $\omega_{a}$ and $\omega_{b}$ compared to $f(z) = \fr{z}{1+z}$. However, both cases suffer the divergence of the error estimate as $z$ increases. Given the covariance matrix, the error on $\omega$ is given by
\ba \sigma_{\omega} &=& \sum_{l=1}^{n} \Bigl(\fr{\partial \omega}{\partial \omega_l}\Bigr)^2 C_{ll} + 2 \sum_{l=1}^{n} \sum_{m=l+1}^{n} \Bigl(\fr{\partial \omega}{\partial \omega_l}\Bigr) \Bigl(\fr{\partial \omega}{\partial \omega_m}\Bigr) C_{lm} \label{sigmaomega} \\ &=& C_{aa} + C_{bb} f(z)^2 + 2 C_{ab} f(z) \, , \nonumber \ea where we use the two parameters models in the second equality. Thus, even though both the linear and the logarithmic parametrization of $\omega$ give the stronger constraints on parameters $\omega_{a}$ and $\omega_{b}$ as shown in the above, both approaches suffer from the instability of the error estimate as $z$ increases. We propose the new parametrization of $\omega = \omega_a + \omega_b \tanh[\ln[1+z]] = \omega_a + \omega_b \fr{2(2+z)}{1+(1+z)^2}$ which is free from the divergence. Then the determinant of the approximate Fisher matrix is given by \be \det(\Fapp)_{H} = \sum_{i=1}^{N-1} \sum_{j=i+1}^{N} G_{i}^2 G_{j}^2 \Bigl( \ln[1+z_i] \ln \Bigl[\fr{1+(1+z_j)^2}{2(1+z_j)} \Bigr] -\ln[1+z_j] \ln \Bigl[\fr{1+(1+z_i)^2}{2(1+z_i)} \Bigr] \Bigr)^2 \, . \label{detFHtanh} \ee This gives the larger $\det(\Fapp)$ compared to CPL.

\begin{figure}
\centering
\vspace{1.5cm}
\begin{tabular}{cccc}
\epsfig{file=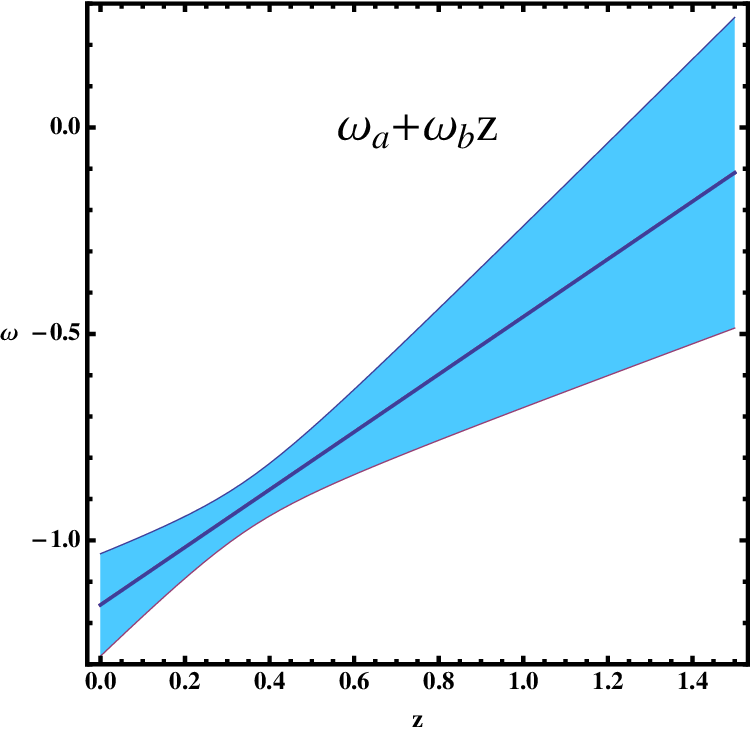,width=0.25\linewidth,clip=} &
\epsfig{file=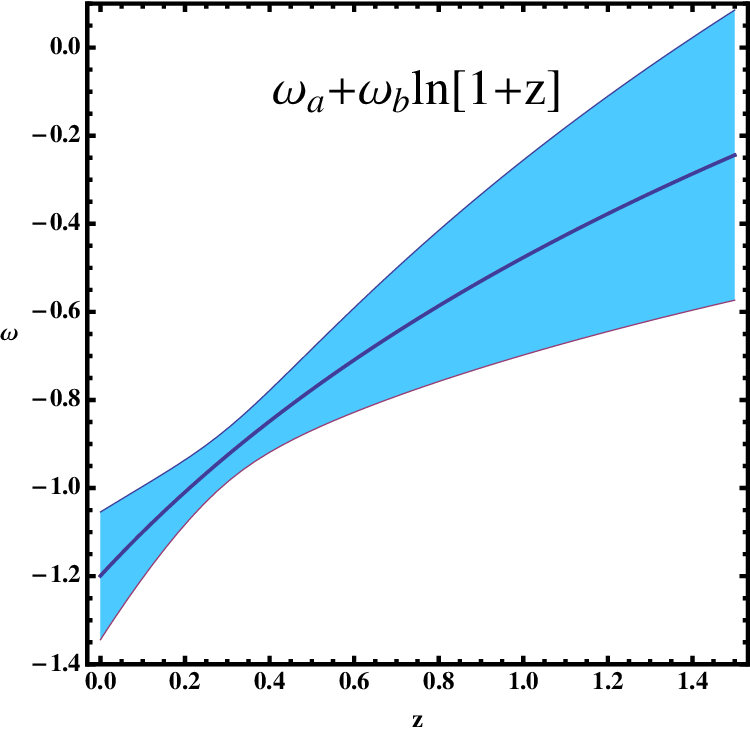,width=0.25\linewidth,clip=} &
\epsfig{file=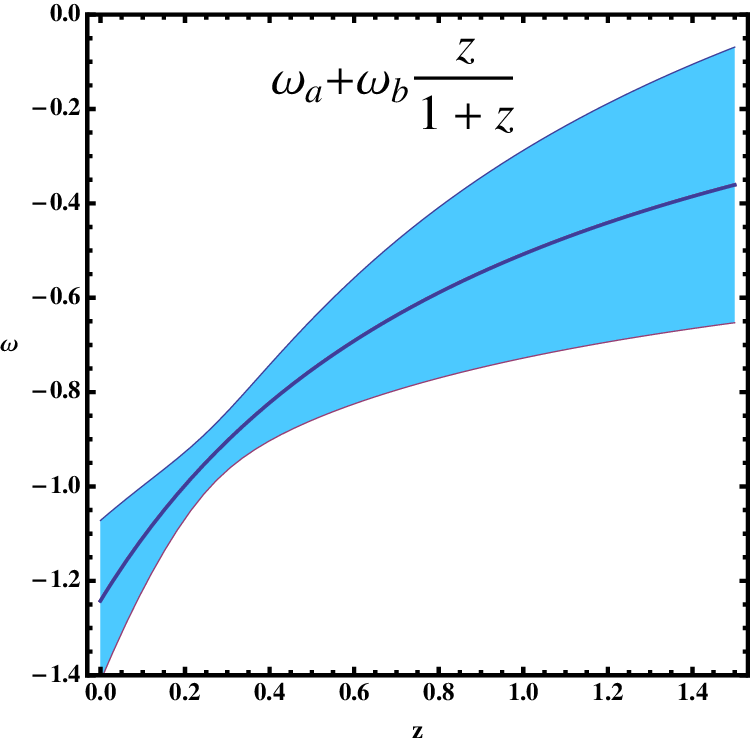,width=0.25\linewidth,clip=} &
\epsfig{file=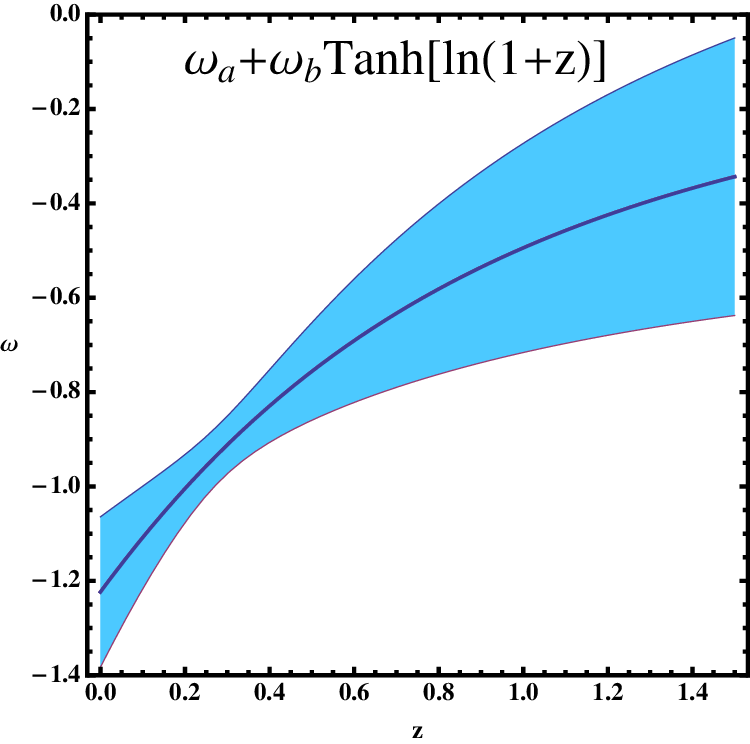,width=0.25\linewidth,clip=}\\
\epsfig{file=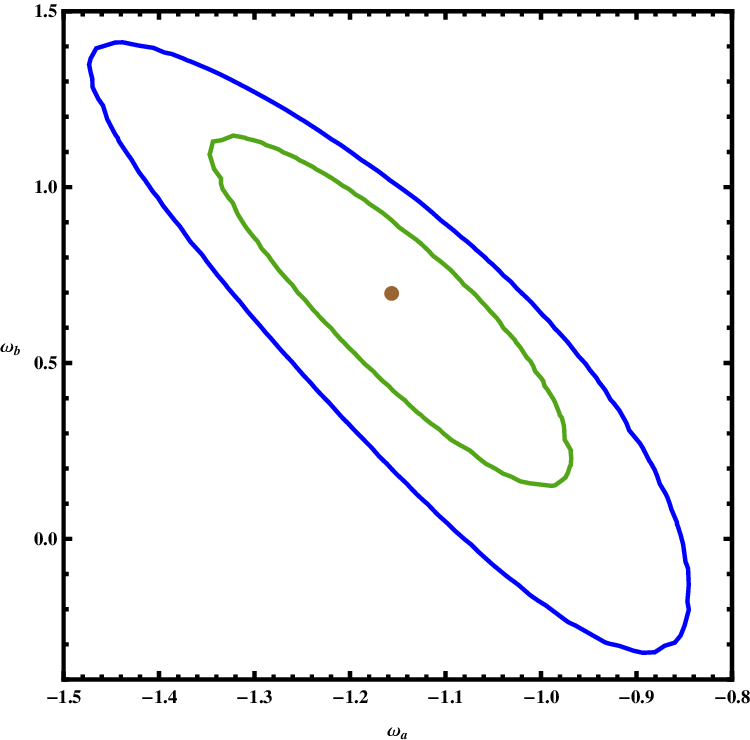,width=0.25\linewidth,clip=} &
\epsfig{file=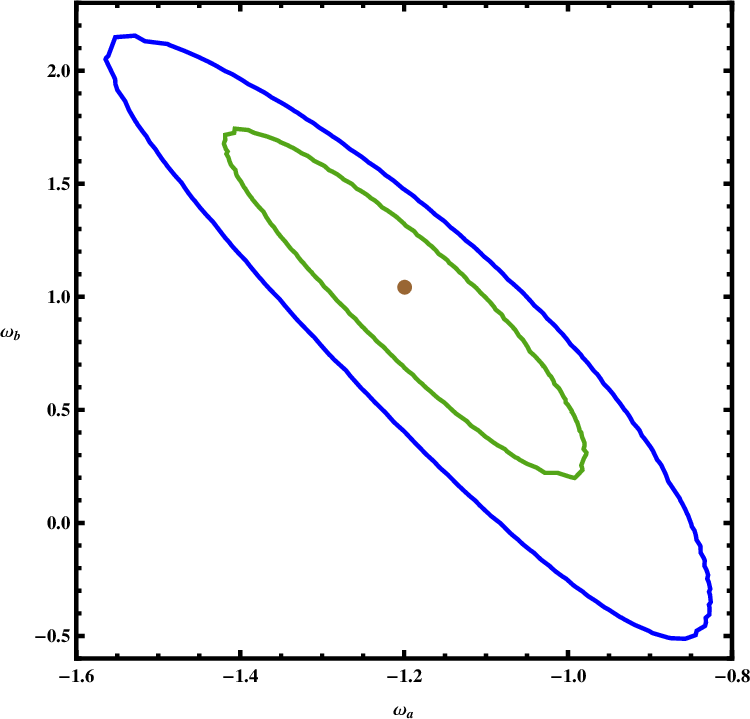,width=0.25\linewidth,clip=} &
\epsfig{file=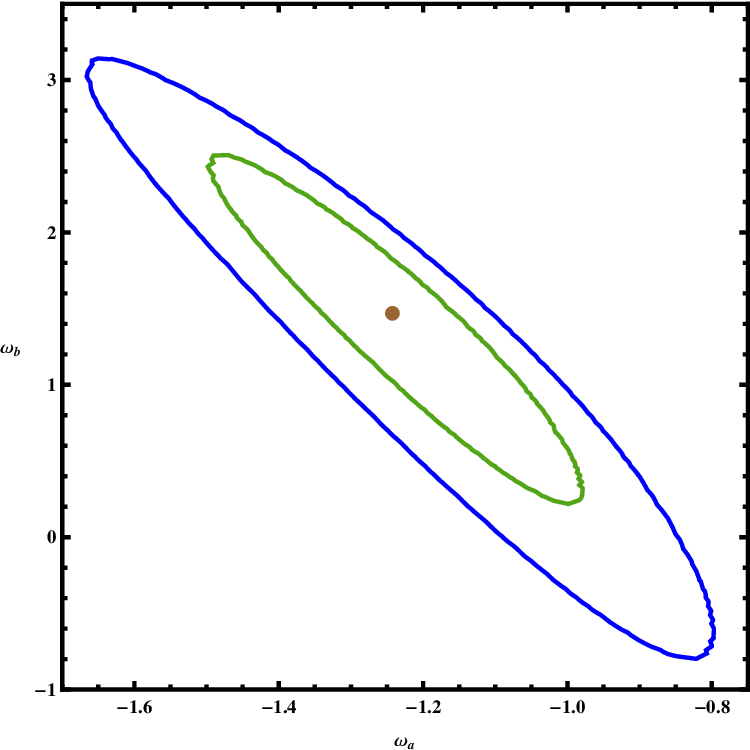,width=0.25\linewidth,clip=} &
\epsfig{file=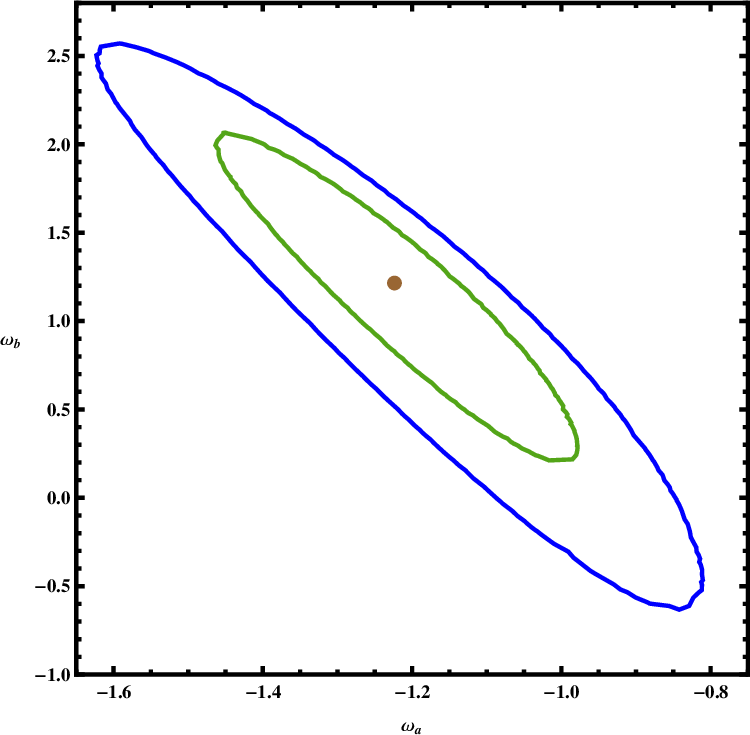,width=0.25\linewidth,clip=} \\
\end{tabular}
\vspace{-0.5cm}
\caption{ a) The variation of the $\w$ over redshift for the different parametrizations. The thick solid lines are the best-fit and the shadow contour represents the $1$-$\sigma$ confidence level around the best fit. b) The $1$-$\sigma$ (inner ellipsoid) and $2$-$\sigma$ (outer one) confidence contours in the $\oa$-$\ob$ plane for the corresponding parametrizations to a).} \label{fig1}
\end{figure}

We show the analysis results for the different parametrizations in table \ref{table2}. We use $H_0 = 71$ km/Mpc/sec and $\Omo = 0.24$ in the analysis. First, we confirm that $ \sum \fr{\Ob_i - \Ob(z_i, \vec{p})}{\sigma_i^2} \fr{\partial^2 \Ob(z_i, \vec{p})}{\partial \omega_a \partial \omega_b}$ is negligible compared to $\sum \fr{1}{\sigma_i^2} \fr{\partial \Ob(z_i, \vec{p})}{\partial \omega_{a}} \fr{\partial \Ob(z_i, \vec{p})}{\partial \omega_{b}}$ in all of models. We also show the best fit values of $\oa$ and $\ob$ for the different parametrizations. The values of the minimum $\chi^2$ for the different models are almost same. But the values of determinant of Fisher matrix for the different models are quite different from one another. As we show in Eq. (\ref{detFH}), linear parametrization produces the largest $\det(F)$ and the so-called CPL parametrization ({\it i.e.} $f(z) = \fr{z}{1+z}$) gives the smallest one among the models.
\begin{center}
    \begin{table}
    \begin{tabular}{ | c | c | c | c | c | c | c | c | c | c | c | c |}
    \hline
      $f(z)$& \multicolumn{3}{|c|}{$ \sum \fr{\Ob_i - \Ob(z_i, \vec{p})}{\sigma_i^2} \fr{\partial^2 \Ob(z_i, \vec{p})}{\partial \omega_a \partial \omega_b}$} & \multicolumn{3}{|c|}{$\sum \fr{1}{\sigma_i^2} \fr{\partial \Ob(z_i, \vec{p})}{\partial \omega_{a}} \fr{\partial \Ob(z_i, \vec{p})}{\partial \omega_{b}}$} & $\omega_{a}^{\ast}$ & $\omega_{b}^{\ast}$ & $\chi^2_{{\rm min}}$ & $\det(F)$ & Ref  \\ \cline{2-7}
          & $(a,a)$ & $(a,b)$ & $(b,b)$ & $(a,a)$ & $(a,b)$ & $(b,b)$ & & & & & \\ \hline
      $z$               & 0.00  & 0.61 & 0.17 & 273 & 93 & 41 & -1.16 & 0.70 & 23.5 & 2604 & \cite{0411803} \\ \hline
      $\ln[1+z]$    & 0.35 & 0.85 & 0.47 & 273 & 73 & 24 & -1.20 & 1.04 & 23.6 & 1115 & \cite{9904356} \\ \hline
      $\fr{z}{1+z}$ & 1.09 & 1.00 & 0.46 & 272 & 59 & 15  & -1.24 & 1.47 & 23.9 & 517 & \cite{0009008, 0208512} \\ \hline
      $\tanh[\ln[1+z]]$ & 0.78 & 1.05 & 0.58 & 272 & 68 & 20 & -1.22 & 1.22 & 23.8 & 783 & \\ \hline
    \end{tabular}
    \caption{Comparison of analysis results for the linear, logarithmic, CPL, and $\tanh[\ln[1+z]]$ parametrizations by using $H(z)$ data given in Ref. \cite{09073149}.  The components of the first term of Fisher matrix is negligible compared to the second one for all the models. $\oa^{\ast}$ and $\ob^{\ast}$ are the best fit values of $\oa$ and $\ob$, respectively. $\chi^2_{{\rm min}}$ are almost identical to each other. We use $H_0 = 71$ km/Mpc/sec and $\Omo = 0.24$.}
    \label{table2}
    \end{table}
\end{center}
Again, we show the above results in Fig. (\ref{fig1}). In each column of the first row, we show the evolution of $\omega$ for the different parametrizatons. The thick solid lines are the best fit $\omega$ and the shadow contour represents the $1$-$\sigma$ confidence level around the best fit. The linear and the logarithmic parametrizations diverge as $z$ increases. Both CPL and $\tanh[\ln[1+z]]$ are free from this divergence and both are almost identical. In the second row, we show the $1$-$\sigma$ (inner ellipsoid) and $2$-$\sigma$ (outer one) confidence contours in the $\oa$-$\ob$ plane for the corresponding parametrizations. As we explain, the linear parametrization gives the strongest constraints on $\oa$ and $\ob$ and the CPL produces the weakest constraints among the models.

\begin{figure}
\centering
\vspace{1.5cm}
\begin{tabular}{ccc}
\epsfig{file=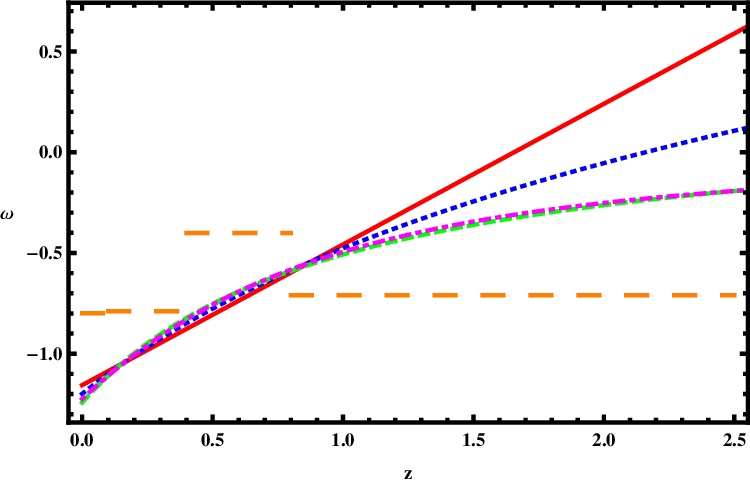,width=0.3\linewidth,clip=} &
\epsfig{file=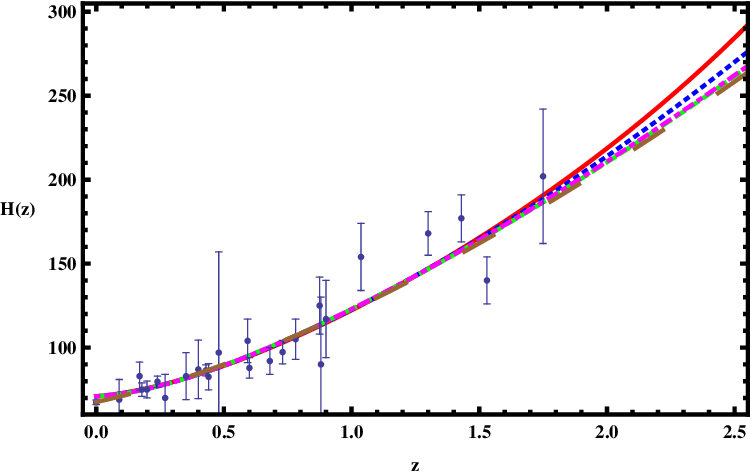,width=0.3\linewidth,clip=} &
\epsfig{file=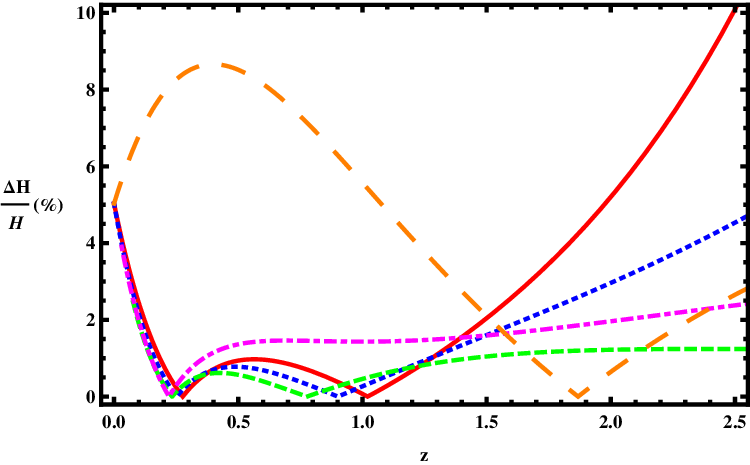,width=0.3\linewidth,clip=} \\
\end{tabular}
\vspace{-0.5cm}
\caption{ a) The evolution behavior of $\omega$ for the different parametrizations b) H(z) obtained from the different parametrizations c) Relative errors of H(z) between each model and  $\Lambda$CDM.} \label{fig2}
\end{figure}

We repeat the analysis by using the so-called the principal component analysis (PCA). For the given $H(z)$ data, we obtain $\oa = -0.799$, $\ob = -0.789$, $\omega_{c} = -0.401$, and $\omega_{d} = -0.710$ if we choose the $4$ principal components at $z_{a} = 0.1$, $z_{b} = 0.4$, $z_c = 0.8$, and $z_d = 1.25$. We emphasize that $\omega_{d}$ is induced values from the other three $\omega_{i}$s as expalined in Ref. \cite{10051770, 11041137}. If one does not use the constraints on $\omega_d$, then $H(z)$ shows the discrepancy from the true value. We compare the evolutions of $\omega$ and $H(z)$ obtained from each model in Fig. \ref{fig2}. In the first column, we show the evolution behavior of $\omega$ for linear (solid), logarithmic (dotted), CPL (dashed), $\tanh$ (dot-dashed), and PCA (long dashed), respectively. Both CPL and $\tanh$ shows the almost identical behavior for $\omega$. We also show the evolution of $H(z)$ including the data points in the second column. All of $H(z)$ are too close to be distinguished. Thus, we compare the $H(z)$ for the different models with one from $\Lambda$CDM model in the third column. The best fit values of $H_0$ and $\Omo$ in $\Lambda$CDM are $67.6$ km/Mpc/sec and $0.325$, respectively with $\chi_{{\rm min}}^2 = 18.01$. We use $H_0 = 71$ km/Mpc/sec and $\Omo = 0.24$ in other two parameters models. $\Delta H$ means the difference of $H(z)$ obtained from the parametrization and the $\Lambda$CDM. Except PCA, errors are smaller than $5$ \% for $z \leq 2$. The fit uses the data up to $z = 1.75$ and that is why there exists bigger error at $z > 2$ for each model.

We can extend the previous consideration to models with $n$ parameters and $N$ datapoint. Then, the determinant of the approximate Fisher matrix is obtained from Eq. (\ref{DetF}) \ba \det(\Fapp) = \sum_{i=1}^{N-n+1} \sum_{j=i+1}^{N-n+2} \cdots \sum_{l=n}^{N} G_{i}^2 G_{j}^2 \cdots G_{l}^{2} \Biggl( \varepsilon_{ij \cdots l} W_{a}(z_i) W_{b}(z_j) \cdots W_{n}(z_{l}) \Biggr)^2 \, . \label{detFn} \ea Thus, the determinant of the $\Fapp$ is composed of $\binom{N}{n}$ number of components with $G_{i}^2 G_{j}^2 \cdots G_{l}^{2} \Biggl( \varepsilon_{ij \cdots l} W_{a}(z_i) W_{b}(z_j) \cdots W_{n}(z_{l}) \Biggr)^2$ where the binomial coefficient is given by  $\binom{N}{n} = N(N-1)(N-2) \cdots (N-n+1)/n!$. Each component has the $n!$ number of permutations. If we consider the observable $H(z)$, then the number of components inside the bracket of PCA method is smaller than $n!$ depending on the location of the principal components and the redshift distribution of the data. It is due to the fact that $W_{\alpha}(z)$ where $\alpha = a, \cdot, n$ is zero if data point is out of the corresponding principal component bin. Thus, PCA method gives less constraints to estimates than other parametrization for the same number of parameters. We find two important features from this. First, PCA method has the weaker constraints than other methods because it produces smaller $\det(\Fapp)_H$. Second, $\det(\Fapp)$ is sensitive to both the location of the PCA components and the data distribution . It is partly agree to the argument in Ref. \cite{09053383}.

We show the importance of parametrization to have the better constrain on the dark energy. Even for the same data, we can improve the constraints of the dark energy if we find the proper parametrization of it. One can speculate the better parametrization based on this approach. Even though PCA is one of the most model independent parametrization, it suffers from the weaker constrain power than other methods.

\section*{Acknowledgments}
We thanks to the anonymous referee for the useful comments.

\end{document}